# Learning Optimal Control of Water Distribution Networks through Sequential Model-based Optimization

Antonio Candelieri[1][0000-0003-1431-576X], Bruno Galuzzi[2][0000-0002-8518-5352], Ilaria Giordani[2][0000-0002-6065-0473] and Francesco Archetti[2][0000-0003-1131-3830]

[1] Department of Economics, Management and Statistics, University of Milano-Bicocca, 20126, Milan, Italy
[2] Department of Computer Science, Systems and Communication, University of Milano-Bicocca, 20126, Milan, Italy
`antonio.candelieri@unimib.it`

**Abstract.** Sequential Model-based Bayesian Optimization has been successfully applied to several application domains, characterized by complex search spaces, such as Automated Machine Learning and Neural Architecture Search. This paper focuses on optimal control problems, proposing a Sequential Model-based Bayesian Optimization framework to learn optimal control strategies. A quite general formalization of the problem is provided, along with a specific instance related to optimization of pumping operations in an urban Water Distribution Network. Relevant results on a real-life Water Distribution Network are reported, comparing different possible choices for the proposed framework.

## 1 Introduction

Sequential Model-based Bayesian Optimization (SMBO) is a sample-efficient strategy for global optimization (GO) of black-box, expensive and multi-extremal functions [1], where the solution of the problem is traditionally constrained to over a box-bounded search space $X$:

$$\min_{x \in X \subset \mathbb{R}^d} f(x) \qquad (1)$$

SMBO has been successfully applied in several domains, ranging from design problems (new materials, drugs, software, structural design) to robotics, control and finance (a brief overview about application domains is provided in Chap. 7 of [2]).

In the Machine Learning (ML) community, it recently became the standard strategy for Automated Machine Learning (AutoML) [3] and Neural Architecture Search (NAS) [4], which are usually characterized by a search space more complex than a box-bounded domain. More precisely, $x$ can consists of mixed (continuous, integer, categorical) components as well as "conditional", where conditional means that the value of a component $x_{[i]}$ depends on the value of another component $x_{[j]}$, with $i \neq j$.

An example of complex search space in AutoML is related to the optimization of ML pipelines, such as that presented in [5].

Starting from the SMBO advances in the ML domain, we consider in this paper an optimal control problem sharing many characteristics with AutoML. More precisely, we addressed the optimal definition of control rules regulating the ON/OFF switching of pumps in a Water Distribution Network (WDN). The objective is the minimization of the energy-related costs while guaranteeing the supply of the water demand.

As better detailed in Section 3, we provide a mathematical formalization of the problem, where the objective function is black-box, explicit constraints on decision variables make the search-space both complex (i.e., analogously to AutoML, decision variables are discrete and conditional) and, partially, black-box.

The rest of the paper is organized as follows: in Section 2, the methodological background about SMBO and optimization of operations in WDNs is presented. Section 3 provides the mathematical formulation of the optimal control problem considered, along with the proposed solution. Section 4 defines the experimental setting and Section 5 summarizes the results obtained. Finally, conclusions and discussion on advantages and limitation of the proposed approach are provided.

## 2    Background

### 2.1    Sequential Model-based Bayesian Optimization

To solve problem (1), SMBO uses two key components: a *probabilistic surrogate model* of $f(x)$, sequentially updated with respect to new function evaluations, and an *acquisition function* (aka *infill criterion* or *utility function*), driving the choice of the next promising point $x$ where to evaluate $f(x)$ while dealing with the exploitation-exploration dilemma. A typical choice for the probabilistic surrogate model is a Gaussian Process (GP) [6] (in this case, SMBO is also known as GP-based optimization or Bayesian Optimization [7,2]). An alternative probabilistic surrogate model is a Random Forest (RF) [8], an ensemble learning approach which, by construction, can deal with mixed and conditional components of $x$, making RFs more well-suited than GPs to solve problems with these characteristics.

The probabilistic surrogate model – whichever it is – should provide an estimate of $f(x)$ along with a measure of uncertainty about such an estimate, with $x \in X$. These two elements are usually the mean and standard deviation of the prediction provided by the probabilistic surrogate model, denoted by $\mu(x)$ and $\sigma(x)$, respectively.

With respect to the acquisition function, several alternatives have been proposed, implementing different mechanisms to balance exploitation and exploration (i.e., $\mu(x)$ and $\sigma(x)$, respectively) [7,2]. In this paper we focused on a subset of acquisition functions, reported in the experimental setting section.

Due to the sequential nature of SMBO, at a generic iteration $n$ we can denote the set of function evaluations performed so far by $D_{1:n} = \left\{\left(x^{(i)}, y^{(i)}\right)\right\}_{i=1,..,n}$, where $y^{(i)} = f(x^{(i)}) + \varepsilon$, and $\varepsilon \sim \mathcal{N}(\mu_\varepsilon, \sigma_\varepsilon)$ in the case of a noisy objective function.

The probabilistic surrogate model is learned at every iteration, providing the corresponding $\mu^{(n)}(x)$ and $\sigma^{(n)}(x)$. The next promising point, $x^{(n+1)}$, is chosen by solving the auxiliary problem:

$$x^{(n+1)} = \underset{x \in X \subset \mathbb{R}^d}{\operatorname{argmax}} \alpha^{(n)}(x) \qquad (2)$$

where $\alpha^{(n)}(x)$ is the acquisition function, typically $\alpha^{(n)}(x, \mu^{(n)}(x), \sigma^{(n)}(x))$. This auxiliary problem is usually less expensive than the original one, and can be solved by gradient-based methods (e.g., L-BFGS) – in the case that the analytical form of $\mu^{(n)}(x)$ and $\sigma^{(n)}(x)$ is given (i.e., when a GP is used as probabilistic surrogate model) – or GO approaches (e.g., DIRECT, Random Search, evolutionary meta-heuristics, etc.) – in the case that $\mu^{(n)}(x)$ and $\sigma^{(n)}(x)$ are also black-box (i.e., when a RF is used as a probabilistic surrogate model).

Then, the objective function is evaluated at $x^{(n+1)}$, leading to the observation of $y^{(n+1)}$ and the update $D_{1:n+1} = D_{1:n+1} \cup \{(x^{(n+1)}, y^{(n+1)})\}$. The process is iterated until some termination criterion is achieved, such as a maximum number of function evaluations has been performed.

## 2.2 Constrained SMBO

The basic SMBO process presented so far, also referred as "vanilla", is not always well suited to solve real life optimization problems. One of the most relevant characteristics to consider – in the so called "exotic" SMBO – is related to the presence of constraints which make the search space more complex than simply box-bounded [7]. In constrained SMBO, the problem (1) can be rewritten as:

$$\begin{aligned} \min_{x \in X \subset \mathbb{R}^d} & f(x) \\ g_i(x) \leq 0 & \quad i = 1, \dots, n_g \end{aligned} \qquad (3)$$

Solving approaches can be categorized depending on the nature of the constraints: they can be known a-priori and given in analytical form or, on the contrary, they are unknown (aka hidden) and black-box. With respect to the first case, several approaches have been proposed in the GO community [9,10,11], while the second case is more related to simulation-optimization and AutoML [12,13,14,15,16].

A further consideration, with respect to unknown constraints, is that the objective function could be not computable in association with the violation of one constraint, leading to global optimization of partially defined functions [17,18]. Recently, a two-stage approach has been proposed in [19], using Support Vector Machine (SVM) to estimate the portion of the box-bounded search space where the objective function is defined (aka computable). In [19] the even more complicated case is considered, where the "feasible" region within the box-bounded search space is implicitly defined by a set of unknown/black-box constraints. In the second stage a constrained Bayesian Optimization task is performed on the estimated feasible region. This paper makes use of this solving approach.

# 3 Problem definition and solution approach

## 3.1 Optimization of operations in Water Distribution Networks

Optimization of WDNs' operations has been an active research field in the last decades. Optimal pump operation, aimed to minimize energy related costs due to pumping water, has been one of the most relevant topics. A systematic review on solutions for WDNs' operations optimization has been recently provided in [20], where approaches for optimal pumps management are categorized into: *(i) explicit control* of pumps by times to operate and *(ii) implicit control* by pumps' pressures, flows or speeds, as well as tanks levels. Although *explicit control* solutions were the most frequently adopted, the optimization problem (also known as Pump Scheduling Optimization, PSO) could be characterized by a huge number of decision variables in the case that the WDN has many pumps and/or times to operate (e.g., decisions about pump activation every hour on a daily horizon).

Most of the *explicit control* solutions proposed use meta-heuristics, mainly evolutionary strategies, such as in [21,22,23]. However, contrary to SMBO, these strategies are not sample efficient, requiring a huge number of hydraulic simulation runs to identify an optimal pump schedule. More recently, an SMBO approach to PSO has been initially proposed in [24] and then extended in [15] to include unknown constraints on the hydraulic feasibility of the pump schedules proposed by SMBO.

On the other hand, *implicit control* strategies allow to reduce the number of decision variables, but make more complex the search space, due to the introduction of further constraints on and conditions among decision variables. Another important advantage offered by implicit control solutions is that they do not require to specify any time to operate; they usually work by applying simple (control) rules depending on the values of collected measurements. Thus, time to operate is given by the data acquisition rate instead of prefixed timestamps as in *explicit control* solutions.

## 3.2 Learning optimal control rules as a black-box optimization problem

We consider the case of an *implicit control* solution, where pumps are controlled depending on the associated pressure values. In the simplest case, control for a given pump is defined by two different thresholds, $x_{[1]}$ and $x_{[2]}$, and the following rule:

```
IF ( pump's pressure < x[1] AND pump is OFF )
   THEN pump is switched ON
ELSE
   IF ( pump's pressure > x[2] AND pump is ON )
      THEN pump is switched OFF
```

This means that the pump is activated if its pressure is lower than a minimum threshold, $x_{[1]}$, it is deactivated if its pressure exceeds a maximum threshold, $x_{[2]}$, and remains in the current status (ON/OFF) otherwise. Clearly, $x_{[1]}$ and $x_{[2]}$ are the deci-sion variables to optimize with respect to the minimization of energy cost, constrained to water demand satisfaction. A graphical representation of this kind of simple control for a single pump is reported in Figure 1.

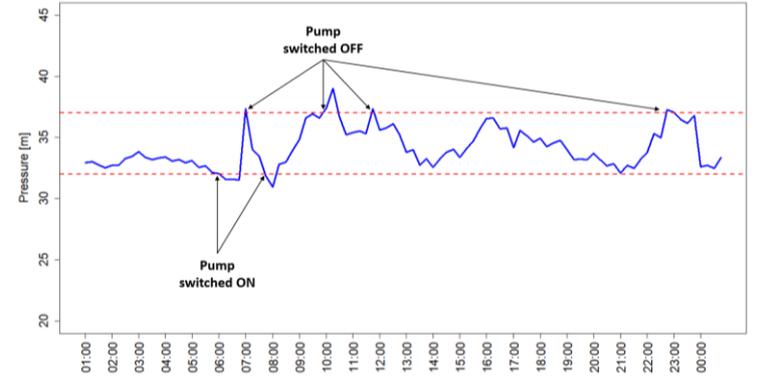

**Fig. 1.** A schematic representation of implicit pump control based on thresholds (red dotted lines) on pressure (in blue). If pressure goes below/over the lower/upper threshold the pump is switched ON/OFF, respectively. Pressure value could not change immediately after the pump switch because it also depends on the status of the other pumps in the WDN

It is important to highlight that both energy costs and water demand satisfaction (as well as any other relevant constraints related to the hydraulic behavior of the WDN, such as min/max tanks levels) are black-box, because they can be only evaluated after having fixed the values of the decision variables. Moreover, an analytical constraint must be added, modelling that the minimum threshold $x_{[1]}$ cannot be greater or equal than the maximum one $x_{[2]}$.

As follows, we define the optimization problem in the more general case consisting of more than a pair of thresholds. This situation is quite common in real-life WDNs, having more than one pump and/or requiring different control thresholds over the day (e.g., during morning and evening) for a given pump:

$$\begin{aligned} &\min_{x \in X \subset \mathbb{R}^d} f(x) \\ &x_{[i]} \in S_i \quad i = 1, \dots, 2\tau \quad (c_1) \\ &x_{[j]} - x_{[j+\tau]} \leq 0 \quad j = 1, \dots, \tau \quad (c_2) \\ &g(x) = 0 \quad (c_3) \end{aligned} \quad (4)$$

where $f(x)$ is the energy cost associated to the control rule defined by the $x_{[i]}$ value, $S_i$ is the set of possible values for the thresholds, $S_i = \{s_1, \dots, s_{N_j}\}$, $\tau$ is the number of thresholds pairs to be set up (leading to $d = 2\tau$) and $g(x)$ is related to the hydraulic feasibility: it is unknown/black-box and makes $f(x)$ partially defined. Thus, both $f(x)$ and $g(x)$ are black-box and are computed via hydraulic software simulation, typically over a simulation horizon of a day. The open-source EPANET 2.0 is the most widely adopted tool for simulating the hydraulic behavior of a pressurized

WDN, so that the search for the optimal values of the control thresholds is sequentially performed on the software model of the WDN before being applied to the real one. A single simulation run, referred to a specific set up of the threshold, involve computational costs; SMBO is a sample-efficient strategy to identify an optimal set up in a limited number of simulation runs (i.e., function evaluations).

Thresholds are modelled as discrete variables in order to consider the resolution of the monitoring sensors (i.e., in the case study analyzed in this paper, measurements are acquired with a resolution of 0.5[m]). In the case of continuous variables, constraint $c_1$ turns into $x_{[i]} \in [\min S_i, \max S_i]$.

According to (4), the optimal definition of an implicit control strategy, based on pressure values, shares common characteristics with AutoML and NAS: decision variables are discrete ($c_1$) and conditional ($c_2$) – such as many Machine and Deep Learning algorithms' hyperparameters – and $g(x)$ is black-box – such as a constraint on resources (i.e., memory usage) for a trained Machine/Deep Learning algorithm.

## 4  Experimental setting

### 4.1  Case study description

The case study considered in this paper refers to a WDN in Milan, Italy, supplying water to three different municipalities: Bresso (around 20'000 inhabitants), Cormano (around 26'000 inhabitants) and Cusano-Milanino (around 19'000 inhabitants). The overall WDN consists of 7418 pipes, 8493 junctions, 14 reservoirs, 1381 valves, 9 pumping stations with 14 pumps overall. Piezometric level of the WDN ranges in 136 to 174 meters (average: 148 meter). Moreover, this WDN is also interconnected with the WDNs of other three municipalities (namely, Paterno Dugnano, Sesto San Giovanni and Cinisello Balsamo). The hydraulic software models of these further municipalities were not available, the hydraulic behavior at the interconnections was modelled through three reservoirs with levels varying over time according to historical data about the flow from the WDN to the other three municipalities and vice versa.

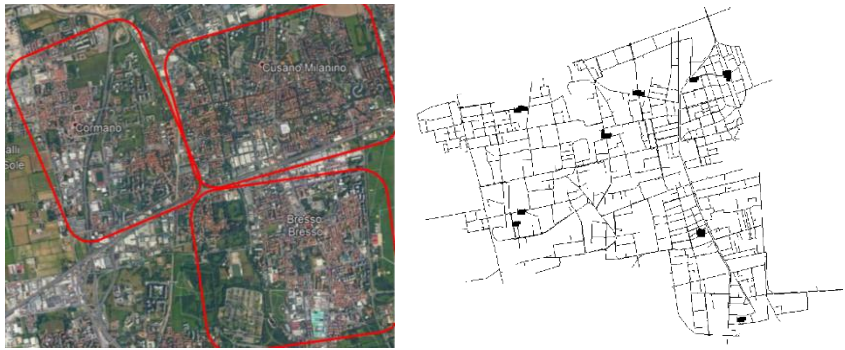

**Fig. 2.** The three municipalities considered in the study (on the left) and the hydraulic software model, developed in EPANET, of the associated WDN (on the right)

### 4.2 SMBO setting

In this section we provide all the details about the setting of our experiments, organized in two different sub-sections. The first one provides all the details about the SMBO process applied to problem (4). In the second, we decided to relax the constraint related to the discreteness of the decision variables. The aim is to evaluate which could be the difference between the optimal solution identified for the problem (4) and a more optimistic one, in the hypothetical case that the numerical precision of the actual control could be finer. In both cases, RF have been used as probabilistic surrogate model due to the presence of conditional decision variables.

Since the actual global optimizer $x^*$ is unknown, we cannot use performance measures such as regret [25] or Gap metrics [26], but just looking at the best value observed over SMBO iterations, the so called "best seen":

$$y^{+\,(n)} = \max_{i=1,\ldots,n}\{f(x^1), \ldots, f(x^t)\}$$

Finally, it is important to highlight that, in this study, we have evaluated all the control strategies identified through SMBO by simulating them over the same "test day". This means that we have considered an unnoisy setting, so $y^n = f(x^{(n)})$, for every $n = 1, \ldots, N$ and with $N$ the maximum number of function evaluations.

#### 4.2.1 RF-based SMBO

As mentioned in Section 4.1, the WDN has 14 pumps, overall. However, 4 are only used to support supply during peak-hours. They are controlled by time and will not be part of the optimization. With respect to the other 10 pumps, 8 of them requires the identification of optimal control thresholds which can be different during the day (i.e., 06:00-23:00) and the night (i.e., 23:00-06:00). This means that we have optimized 2 thresholds for 2 pumps and 4 thresholds for 8 pumps, leading to 36 decision variables overall (i.e., thresholds $x_{[i]}$) for the problem (4) – that is $\tau = 18$. It is important to highlight that the number of decision variables should be significantly higher in the case of *explicit control*: the optimization of hourly-based schedules on the same case study would require 240 decision variables (that is 10 pumps time 24 hours).

The possible discrete values for all the *lower* thresholds, that are the sets $S_{i=1,\ldots,\tau}$, range from 21[m] to 32[m], with a step of 0.5[m] (i.e., 23 possible values). The possible discrete values for all the *upper* thresholds, that are the sets $S_{i=\tau+1,\ldots,2\tau}$, range from 26[m] to 44[m], with a step of 0.5[m] (i.e., 23 possible values). These two sets instantiate the constraint $(c_1)$ of the problem (4).

Initialization of the probabilistic surrogate model (i.e., a RF) was performed by randomly sampling 10 initial vectors of control thresholds ("initial design"). More precisely, a Latin Hypercube Sampling (LHS) procedure has been applied. Remaining budget (i.e., function evaluations) has been set to 200.

We decided to compare three different acquisition functions, namely Lower Confidence Bound (LCB), Expected Improvement (EI) [7,2] and Augmented Expected Improvement (AEI) [27] – the last usually replaces EI in the noisy setting.

$$LCB(x) = \mu(x)^{(n)} - \beta^{(n)}\sigma(x)^{(n)}$$

$$EI(x) = \begin{cases} (y^+ - \mu(x)^{(n)})\Phi(Z) + \sigma(x)^{(n)}\phi(Z) & \text{if } \sigma(x)^{(n)} > 0 \\ 0 & \text{otherwise} \end{cases}$$

$$AEI(x) = \begin{cases} (y^+ - \mu(x)^{(n)})\Phi(Z) + \sigma(x)^{(n)}\phi(Z)\left(1 - \frac{\sigma_\varepsilon}{\sqrt{\sigma_\varepsilon^2 + (\sigma(x)^{(n)})^2}}\right) & \text{if } \sigma(x)^{(n)} > 0 \\ 0 & \text{otherwise} \end{cases}$$

where $\beta^{(n)}$ is used to manage the exploitation-exploration trade-off, $Z = \frac{y^+ - \mu(x)^{(n)}}{\sigma(x)^{(n)}}$ and $y^+$ is the "best seen" up to $n$. In the case of LCB, the next promising point will be $x^{(n+1)} = \underset{x \in X \subset \mathbb{R}^d}{\operatorname{argmin}} LCB(x)$; in the case of EI and AEI it will be given by $x^{(n+1)} = \underset{x \in X \subset \mathbb{R}^d}{\operatorname{argmax}} EI(x)$ and $x^{(n+1)} = \underset{x \in X \subset \mathbb{R}^d}{\operatorname{argmax}} AEI(x)$, respectively.

Since we are using a RF as probabilistic surrogate model, the acquisition functions are also black-box. A global-local method has been used to solve the auxiliary problem (2) and identify the next promising $x^{(n+1)}$. More precisely, the global-local method used is known as "focus-search" [28]: it can handle with numeric, discrete and mixed search spaces, also involving conditional variables. Focus-search starts with a large set of random points where the acquisition function is evaluated. Then, it shrinks the search space around the current best point and perform a new random sampling of points within the "focused space". The shrinkage operation is iteratively performed until a maximum number of iterations and the entire procedure can be restarted multiple times to mitigate the risk to converge to a local optimum. Finally, the best point over all restarts and iterations is returned as the solution of the auxiliary problem (2).

Although different acquisition functions have been used, all the associated SMBO processes started from the same initial design. Furthermore, to mitigate the effect of randomness due to the initial design, we performed 20 different experiments with 20 different initial designs.

### 4.2.2 RF based SMBO with relaxation of the discreteness constraint

In this experiment we have decided to relax the problem (4) by removing the constraint about the discreteness of the decision variable ($c_1$). This makes the initial box-bounded search space continuous, even if it remains complex due both to the presence of conditional decision variables ($c_2$) and the black-box constraint related to the feasibility of the hydraulic simulation ($c_3$). The rest of the experimental setup is identical to what reported in the previous sub-section.

## 5 Results

This section summarizes the most relevant results. Figure 3 shows how the "best seen" changes over function evaluations: solid lines are the averages over 20 different runs, while the shaded areas represent the standard deviations (almost 0). The first value, at iteration 0, is the best seen observed within the initial design.

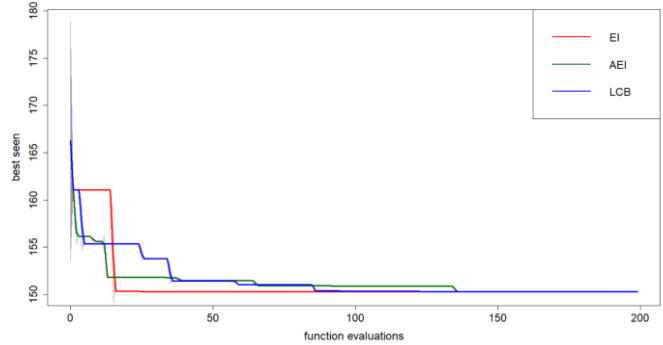

**Fig. 3.** Best seen over function evaluations of RF-based SMBO using three acquisition functions: EI (red), AEI (green) and LCB (blue). Solid lines and shaded areas represent, respectively, mean and standard deviation (that is almost 0) of the best seen over 20 different runs

With respect to the second experiment – related to the relaxation of the discreteness constraint ($c_1$) of problem (4) – Figure 4 shows how the "best seen" changes over function evaluations. Visualization has been limited to the first 20 function evaluations – out of the overall 200 – because, already after two function evaluations, no further improvements have been obtained.

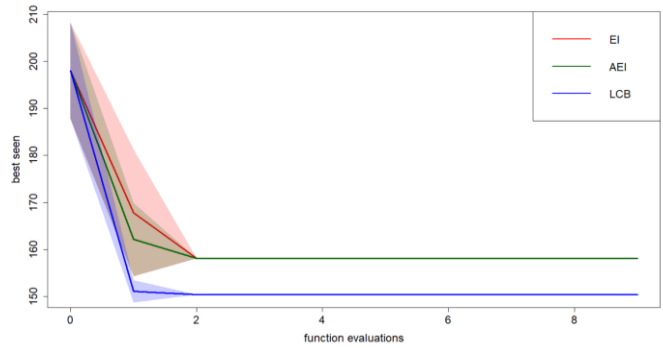

**Fig. 4.** Best seen over function evaluations of RF-based SMBO with relaxation of the discreteness constraint. Comparison between three acquisition functions: EI (red), AEI (green) and LCB (blue). Solid lines and shaded areas represent, respectively, mean and standard deviation (that is almost 0) of the best seen over 20 different runs

Finally, we have evaluated the improvement, in terms of energy costs reduction, provided by SMBO with respect to the energy cost implied by the current pressure-based control operated by the water utility, that is 332,30€/day. In Table 1, the best cost over 20 runs has been selected for every acquisition function and separately for the two types of experiments described in section 4.2.1 and 4.2.2.

**Table 1.** Optimal energy costs obtained via SMBO and associated costs reduction with respect to the current cost implied by the current pressure-based control operated by the water utility

|  | Original Problem (4) | | | Relaxation of ($c_1$) | | |
|---|---|---|---|---|---|---|
|  | EI | AEI | LCB | EI | AEI | LCB |
| Energy cost [€] | _**150.29**_ | 150.32 | 150.31 | 158.12 | 154.91 | 150.42 |
| Cost reduction w.r.t. the currently operated control strategy [€] | _**182.01**_ | 181.98 | 181.99 | 174.18 | 177.39 | 181.88 |

The relaxation of discreteness constraint, $c_1(x)$, does not provide any improvement. This could be due to the use of RF, which can result less effective on continuous variables than discrete ones, also depending on the smoothness of the objective function. Probably, in the second experiment, the approach was not able to escape from some plateau, within the maximum number of function evaluations allowed.

As an example, we report in Figure 5 the activation pattern of two pumps, A and B, according to the control currently operated by the WDN ("curr" suffix) versus the new activation implied by the new control optimized through SMBO ("opt" suffix).

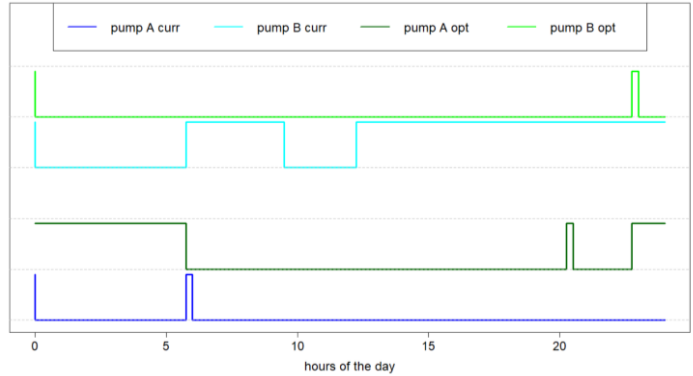

**Fig. 5.** Activation of two pumps according to current and optimized implicit control

## 6     Conclusions and discussion

We have presented a SMBO approach for solving optimal control problems characterized by black-box objective functions and complex, partially unknown, search spaces. A general formalization of the problem was provided along with an instantiation on a specific real-life application, that is the optimal control of pumps in water distribution networks. The use of a hydraulic simulation software, EPANET, makes both objective function and constraints – related to hydraulic feasibility of the identified control rules – black-box.

Using SMBO to search for an optimal *implicit* control allowed us to work with a dimensionality which is significantly lower than the one required by the (more widely adopted) *explicit* controls.

A more realistic experimentation should consider different "simulation days", characterized by random water demands whose empirical distribution is generated from historical data. This requires evaluating the robustness of the implicit control rules proposed by SMBO and to move towards a "distributionally robust" SMBO.

## 7 Acknowledgements

This study has been partially supported by the Italian project "PerFORM WATER 2030" – programme POR (Programma Operativo Regionale) FESR (Fondo Europeo di Sviluppo Regionale) 2014–2020, innovation call "Accordi per la Ricerca e l'Innovazione" ("Agreements for Research and Innovation") of Regione Lombardia, (DGR N. 5245/2016 - AZIONE I.1.B.1.3 – ASSE I POR FESR 2014–2020) – CUP E46D17000120009.
We greatly acknowledge the DEMS Data Science Lab for supporting this work by providing computational resources.